\documentclass[final,5p,times,twocolumn]{elsarticle}

\usepackage{graphicx}       
\usepackage{hyperref}       
\hypersetup{
    colorlinks=true,
    linkcolor=blue,
    citecolor=blue,
    urlcolor=blue,
}
\usepackage{algorithm}
\usepackage{algorithmic}
\usepackage{amsmath}
\usepackage{amssymb}

\usepackage[caption=false]{subfig}

\usepackage{placeins}

\journal{Acta Materialia}  

\begin{document}

\begin{frontmatter}

\title{Towards Autonomous Experimentation: Bayesian Optimization over Problem Formulation Space for Accelerated Alloy Development}

\author[1]{Danial Khatamsaz\corref{cor1}}
\ead{Khatamsaz@tamu.edu}
\author[2]{Joseph Wagner}
\ead{Joewag@tamu.edu}
\author[1]{Brent Vela}
\ead{Brentvela@tamu.edu}
\author[1,2,3]{Raymundo Arr\'oyave}
\ead{Rarroyave@tamu.edu}
\author[2,cor1]{Douglas L. Allaire}
\ead{Dallaire@tamu.edu}

\cortext[cor1]{Corresponding author: dallaire@tamu.edu}
\cortext[cor2]{Corresponding author: rarroyave@tamu.edu}

\address[1]{Department of Materials Science and Engineering, Texas A\&M University, College Station, TX 77843, USA}
\address[2]{J. Mike Walker '66 Department of Mechanical Engineering, Texas A\&M University, College Station, TX 77843, USA}
\address[3]{Wm Michael Barnes '64 Department of Industrial and Systems Engineering, Texas A\&M University, College Station, TX 77843, USA}

\begin{abstract}
Accelerated discovery in materials science demands autonomous systems capable of dynamically formulating and solving design problems. In this work, we introduce a novel framework that leverages Bayesian optimization over a problem formulation space to identify optimal design formulations in line with decision-maker preferences. By mapping various design scenarios to a multi-attribute utility function, our approach enables the system to balance conflicting objectives—such as ductility, yield strength, density, and solidification range—without requiring an exact problem definition at the outset. We demonstrate the efficacy of our method through an \emph{in silico} case study on a Mo-Nb-Ti-V-W alloy system targeted for gas turbine engine blade applications. The framework converges on a "sweet spot" that satisfies critical performance thresholds, illustrating that integrating problem formulation discovery into the autonomous design loop can significantly streamline the experimental process. Future work will incorporate human feedback to further enhance the adaptability of the system in real-world experimental settings.

\end{abstract}


\end{frontmatter}

\section{Introduction}

The design process in engineering and science typically follows a series of well-established steps: identifying design objectives, defining constraints, exploring the design space, optimizing solutions, and validating through experimentation \cite{Arroyave2022ADesign}. These steps guide the decision-making process, ensuring that a design meets the desired performance criteria. Traditionally, each step has relied heavily on human intuition and expertise, especially the critical stage of problem formulation, where objectives and constraints are clearly defined based on the designer's goals and preferences.

\subsection{Formalizing the Problem Formulation Challenge}
Among these steps, problem formulation is perhaps the most pivotal, as it establishes the framework within which the entire design campaign unfolds. Traditionally, this task has been considered a domain reserved for human input, since it involves encoding the designer’s preferences, objectives, and trade-offs---an inherently subjective task. It is this formulation that shapes the trajectory of the entire design campaign. The human designer, drawing from experience and domain knowledge, specifies what is to be maximized or minimized, sets priorities for competing objectives, and identifies relevant constraints that ensure feasibility.

However, it is often the case that, despite the input of expert opinion, the design campaign is not properly formulated at the outset. As more experimental data are gathered throughout the campaign, the design objectives and constraints are frequently---and necessarily---modified \cite{acemi2024multi}. This iterative learning process can reveal that the initial formulation was misaligned with the true goals of the discovery effort. At the start of a project, limited data and preliminary assumptions may lead to an oversimplified or even outdated understanding of the problem. Unforeseen complexities and interactions often emerge as the campaign progresses, highlighting that the early objectives did not fully capture the underlying challenges. Moreover, advances in technology and methodology, along with inherent human biases and shifting priorities, can further expose the deficiencies of the original problem statement. Only later in the campaign do experts realize that they have been searching for the wrong material or optimizing for the wrong criteria due to an ill-formed design problem. This evolving understanding underscores the need for a flexible, dynamic approach to problem formulation, one that can adapt to new insights as they arise.

In the specific case of multidisciplinary approaches to closed-loop alloy design campaigns \cite{hastings2024interoperable,mulukutla2024illustrating}, experts from different fields often have conflicting opinions on objectives, constraints, and their relative importance. These disagreements can hinder progress if not addressed, highlighting the need for an impartial arbiter. This arbiter, whether in the form of a decision-making framework or computational tool, must reconcile these differing viewpoints, ensuring that all relevant perspectives are considered while dynamically adjusting to new data and preferences.

Advancements in computational modeling, automation, and experimental techniques have sparked scientific innovation across many disciplines, prompting a reevaluation of traditional design approaches~\cite{shi2021automated}. In response, \emph{autonomous design} has emerged as a promising solution, aiming to integrate the problem formulation stage within the automated experimentation loop. This shift is essential to fully automate the discovery process, enabling the system not only to run experiments but also to dynamically adjust design goals in response to evolving knowledge and the needs of the stakeholders who will ultimately benefit from the designed object, system, or material. The key challenge lies in efficiently leveraging available resources to develop a comprehensive understanding of the system and achieve optimal performance. To meet this challenge, autonomous systems must be capable of formulating the design problem on their own, incorporating the preferences and objectives that were once solely determined by human designers.

\subsection{Toward Autonomous Design and Discovery}
The goal of developing autonomous design or discovery systems---self-driving labs in the context of chemical or materials discovery---is to replace human effort with machines to enable faster and more efficient discoveries~\cite{CRABTREE20202538} across different scientific fields~\cite{ament2021autonomous,bash2021multi,langner2020beyond,gongora2020bayesian,deneault2021toward,epps2020artificial,christensen2021data,grizou2020curious,cao2021optimization}. Artificial intelligence drives autonomous experimentation by interpreting experimental outcomes to propose and refine subsequent tasks~\cite{szymanski2021toward,epps2020artificial}. For instance, self-driving laboratories combine automation and AI to manipulate a vast set of experimental variables~\cite{macleod2022self}. These systems employ a closed-loop approach, learning from past experiments to continuously adjust goals and ensure that the most valuable experiments are executed~\cite{hase2019next}. An example is provided in Ref.~\cite{gongora2020bayesian}, where Bayesian optimization drives a mechanical testing system that automatically 3D prints and tests parts to determine their mechanical properties. In other works, multi-objective Bayesian optimization is implemented in self-driving laboratories to accelerate the discovery of materials with multiple objectives~\cite{macleod2022self,macleod2021advancing}.

A common ground among these studies is that the specific design problem is known \emph{a priori}. That is, the designer's preferences are encoded as the minimization or maximization of different quantities of interest (QoIs) and a set of design constraints at the outset of the campaign, and these remain fixed throughout. Various techniques can be implemented to solve a single design problem formulated by designers. For instance, one may take a multi-objective optimization approach~\cite{deb2011multi,konak2006multi,mingqiang2000ga,nedjah2015evolutionary,khatamsaz2021bayesian,arroyave2022perspective,alvi2024hierarchical} to discover the Pareto front representing all trade-off solutions, thereby allowing the designer to choose a preferred solution. Alternatively, one may create a single scalar value~\cite{knowles2006parego,hakanen2017using} based on weighted objectives and then perform single-objective optimization to find the optimal solution accordingly. Although the discovered solution lies on the Pareto front, its location depends on the importance assigned to each QoI by the designer.

However, as the design process unfolds, the preferences may change, necessitating a reformulation of the problem. This leads to the costly repetition of the design process from scratch. Another significant challenge arises when more than a few QoIs are involved: the limited capacity for human interpretation of high-dimensional data complicates the selection of a single best design among multiple candidates. While encoding the preferences into mathematical expressions to create a value function could, in principle, allow for quantitative comparisons, the complex interactions among multiple QoIs often render such value functions impractical.

In this study, we aim to improve design autonomy by introducing the notion of optimization over problem formulation space. The idea is to let the system iteratively decide on the most valuable problem to solve by searching a problem formulation space created based on all possible design scenarios concerning every QoI. Later, we describe how different design problems can be identified in a problem formulation space by creating a linear combination of QoI indicators that also enables a distance-based metric to correlate problem formulations accordingly.

\subsection{Active-Learning of Optimal Problem Formulation}
AI-enabled active-learning approaches such as Bayesian optimization can be implemented to look for the optimal problem formulation that yields the highest utility to the designer based on their most recent preferences. By taking the designer's preferences as iteratively updated feedback to the system, the framework can explore the space of problems while exploiting information from previously solved ones. Such autonomy facilitates more efficient use of resources while letting the designer update their preferences in an online manner as the design process unfolds and new information is revealed.

\begin{figure*}[htb!]
\centering
\includegraphics[width=\textwidth]{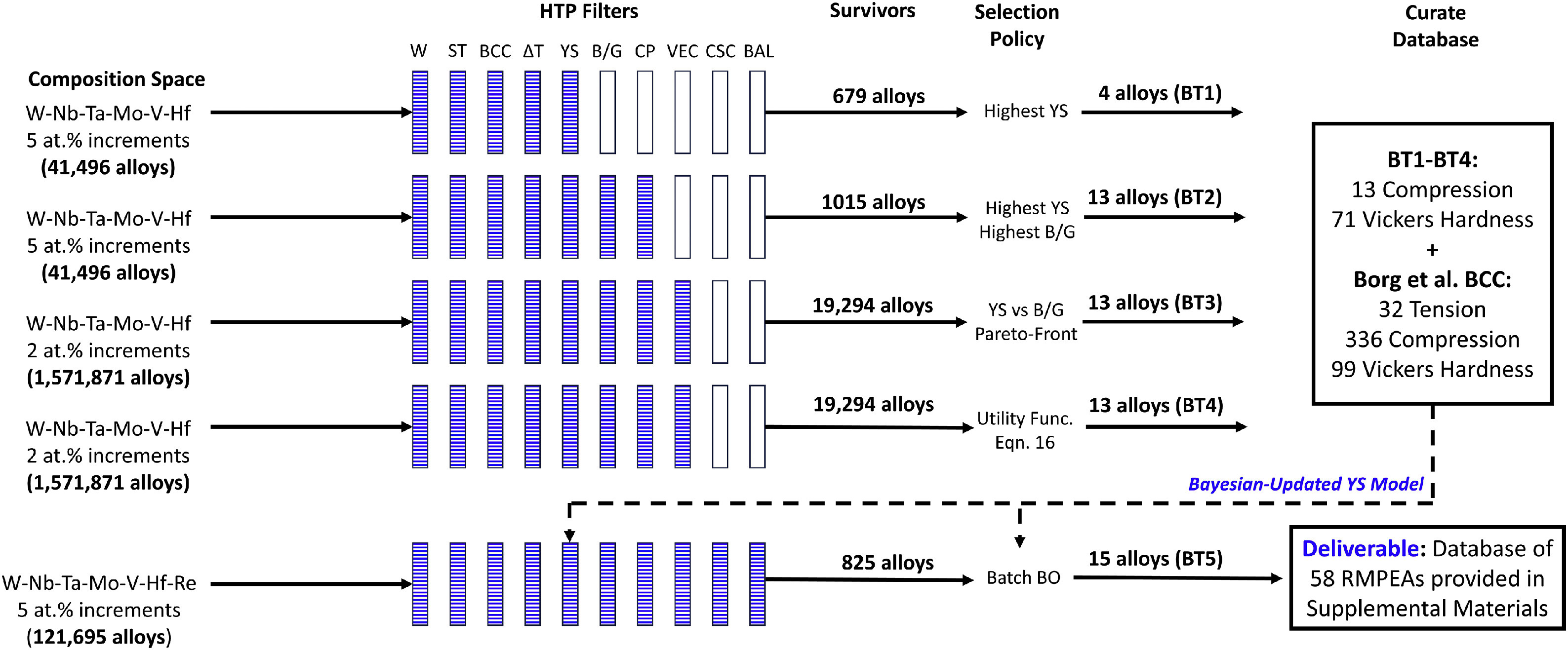}
\caption{Schematic of a real development campaign to identify refractory alloys for ultra-high temperature applications up to 2000 $^{\circ}$C---see Ref. \cite{acemi2024multi} for more details. Rather than a fixed optimization problem, the development effort turned into a highly dynamic process where the alloy design problem formulation changed at each batch (BT): BT1–4 differ by the addition of constraints and changes in the selection policy, informed by experimental results. BT5 is the first batch to use machine learning for “closing the experimental loop” via Bayesian optimization. W: Tungsten Constraint. ST: Solidus temperature constraint. BCC: phase stability constraint. YS: Yield strength constraint. B/G: Pugh ratio ductility constraint. CP: Cauchy pressure ductility constraint. VEC: Valence electron concentration ductility constraint. CSC: Hot cracking susceptibility coefficient constraint. BAL: Balling resistance constraint. Reproduced with permission from \cite{acemi2024multi}.}
\label{esi}
\end{figure*}

Most often, formulating a specific problem at the beginning stages of a design campaign has to be done with little to no information about a target system in the hope of achieving optimal performance metrics \cite{acemi2024multi}. Acquiring more information may necessitate constantly reformulating the problem, which imposes huge loss in terms of cost and time. In contrast to a closed-loop optimization framework that searches for the optimal designs according to the specific problem objectives, our proposed discovery loop searches for the optimal problem to solve that is most likely to satisfy the most recent designer's preferences without constraining the discovery to a particular design task. Thus, the framework directly interacts with preferences instead of objectives.

As mentioned earlier, creating a value function for quantitative comparison of different designs is usually impractical as there are no systematic means to elucidate preferences regarding interactions between multiple QoIs. However, an A/B testing approach enables a more viable way of inputting the designer's preferences into the system. Given a pair of designs and respective properties, the designer can choose the preferred one. By repeating this process and collecting the designer's feedback on solutions to multiple problems, the system exploits the information to decide on the next problem to solve. The discovery loop is driven by Bayesian optimization that employs a probabilistic model to predict a designer's preferences by assigning a value to each problem formulation and computing the expected improvement associated with problem formulation candidates. Such flexibility in the discovery loop enables updating the preferences progressively and moving toward different regions of the problem formulation space as needed without losing any information acquired by solving problems according to outdated designer preferences. This approach proves particularly beneficial in domains where design objectives are inherently complex, such as materials development.

\subsection{Application to Materials Discovery and Development}

Designing materials for specific engineering applications requires achieving optimal performance with respect to multiple properties to satisfy a variety of operational conditions to which a material is exposed~\cite{Arroyave2022ADesign}. Most often, target properties are conflicting, and improving one does not necessarily improve other properties. Here, expert preferences play a key role in determining how trade-offs should be addressed in order to discover the best material possible to satisfy all required performance metrics to a degree. As mentioned earlier, experts may need to reformulate the problem as more information from the system is revealed after new observations are made. Moreover, minimization or maximization of a property is not always the best way to describe the desired performance requirements of a material. In many cases, optimizing a property past a certain threshold not only imposes additional costs but also results in deterioration of conflicting properties. These complicated interactions make it nontrivial to formulate a single optimization problem and build a value function for quantitative comparisons, especially at the early stages of a design campaign. However, an expert may be able to compare two designs and their associated properties and choose the better option without specifically defining a design problem. Our proposed AI-assisted, closed-loop discovery of optimal problems can take expert input in the form of A/B tests between solutions to already solved design problems and decide on the most valuable problem formulation accordingly. The goal is to discover the design that wins the most A/B tests, and best satisfies expert preferences.

Almost all engineering sectors, including aviation, automotive, power generation, and energy storage, are constantly seeking solutions to enhance system efficiency. A key step in achieving this is discovering new materials with superior properties. One of the most pressing engineering challenges is developing materials for turbine blades~\cite{arpa-e_2020}. These materials must possess excellent manufacturability and withstand higher operational temperatures and speeds to improve the thermal efficiency of power generation systems.

Efforts to discover improved alloys for turbine blades have leveraged multi-objective Bayesian optimization approaches to identify the Pareto front associated with multiple important properties~\cite{khatamsaz2022multi,khatamsaz2023bayesian}. In these approaches, objective functions are typically defined as the minimization or maximization of specific properties. Ignoring the thresholds and looking for designs past a minimum required performance (after applying engineering safety factors and such) for turbine blades not only increases the cost associated with the discovery of unnecessary and invaluable designs but also leaves the expert with many different potential solutions without providing a basis to choose the final design. On the other hand, using single-objective Bayesian optimization entails the expert making an early decision on the importance of each property to construct a scalar objective function with little to no information. Consequently, updating preferences along the way requires restarting the design campaign from scratch and losing resources and time.

A recent example of such a campaign has been reported by a subset of the present authors\cite{acemi2024multi} as part of an effort to develop tungsten-based refractory alloys for operations beyond 2,000$^{\circ}C$---see Fig.~\ref{esi}. While initially planned as a well-defined optimization campaign, the alloy development effort became a highly dynamic and evolving program: Initially, basic constraints and objectives were imposed---minimum tungsten content, a single-phase BCC structure from 2000$^{\circ}$C to the solidus, yield strength above 50 MPa at 2000$^{\circ}$C, and a narrow solidification range to mitigate hot cracking. As the project progressed, early parameters proved inadequate. In Batch 2, the solidus requirement was raised, and ductility criteria were added. Batch 3 increased compositional resolution and introduced a valence electron concentration constraint while relaxing the yield strength threshold to 40 MPa to favor ductility. Batch 4 replaced simple filtering with a composite objective function that integrated yield strength, ductility, and solidification behavior. Finally, Batch 5 expanded the design space to include rhenium, incorporated additional printability metrics, and updated the yield strength model via Bayesian optimization to better integrate new data. In the end, several candidate optimal alloys were identified, yet, the trajectory toward optimality was highly dynamic and, in hindsight, not really aligned with conventional iterative discovery approaches such as those based on Bayesian Optimization\cite{arroyave2022perspective}.

To demonstrate our proposed problem discovery framework as part of achieving autonomous design, we present a simple scenario focused on discovering high-entropy alloys (HEAs) for manufacturing turbine blades. HEAs, known for their exceptional characteristics like strength, thermal stability, wear resistance, and hardness, are prime candidates for applications in harsh environments. In this scenario, we explore a 5-element system consisting of Nb-Mo-Ti-V-W. The key quantities of interest (QoIs) for evaluating alloy performance include Cauchy pressure (as an indicator of ductility), yield strength, density, and solidification range.

Here, we define the discovery campaign over a static A/B testing scenario where the preferences do not change over time. In other words, we replace human effort with utility functions from the beginning of the campaign without further updating the utility functions so that the comparisons of alloys always yield the same ranking and utility values. The goal is to maximize a multi-attribute utility function, which combines these individual utility functions based on their relative priority.

\subsection{Overview of the Present Work}
To formalize this problem within the problem discovery framework, we define a structure that allows us to correlate different design problems by measuring point-wise distances, providing the foundation for employing a distance-based kernel function. With this, we can map problem formulations to a utility value using probabilistic modeling---commonly Gaussian processes. Bayesian optimization is then employed to iteratively explore and learn within the problem formulation space, progressively searching for the optimal problem formulation.

In the following section, we first establish the basics to define a problem formulation space and introduce the proposed autonomous design framework by going through a step-by-step description of its ingredients. In Sec.~\ref{demonstration}, we demonstrate the application in the discovery of high-entropy alloys with improved properties to enable more efficient power generation in gas turbines. Discussion and concluding remarks are presented in Sec.~\ref{conclusions}.

\section{Methodology}
\label{Methodology}
\subsection{Problem Formulation}
We define a problem formulation space as:
\begin{equation}
    P_K = \{ p \in \mathcal{P}: p= \min_{\textbf{x}\in \mathcal{X}}\; \mathbf{Q}(\textbf{x}) = 
    \begin{bmatrix}
    Q_1 (\textbf{x})\\
    Q_2 (\textbf{x})\\
    \vdots\\
    Q_K (\textbf{x})\\
    \end{bmatrix}\}
\end{equation}
where a problem can be defined over any subset of $K$ different QoIs.

Later in this section, we demonstrate that any constraints can be incorporated into the formulation as an additional QoI. To address such optimization problems, one may employ multi-objective techniques to approximate the Pareto front of solutions with trade-offs using hypervolume-based methods~\cite{Beume2009S-MetricProblem,Bradstreet2010ACalculations,Emmerich2011Hypervolume-basedComputation,Fonseca2006AnIndicator,Russo2012QuickHypervolume,Yang2007NovelSet,Zitzler1999MultiobjectiveApproach}. Alternatively, one can convert the multi-objective problem into a single-objective one by constructing a scalar objective function using weighted sums, where each QoI is weighted according to the designer's preferences~\cite{Marler2010TheInsights,Kim2005AdaptiveGeneration}. The latter approach is often more practical in real-world applications since obtaining the entire Pareto front is typically cost-prohibitive, and many solutions at the extremes may fall outside the designer’s desired range. By varying the weights, an infinite number of problems can be defined, with a larger weight on a particular QoI guiding the solution closer to its global minimizer. In this framework, adjusting the weights effectively explores all definable problem formulations involving the QoIs.

\subsection{Normal Boundary Intersection Method for Optimal Problem Formulation}

\begin{figure}[htb!]
\centering
\includegraphics[width=\columnwidth]{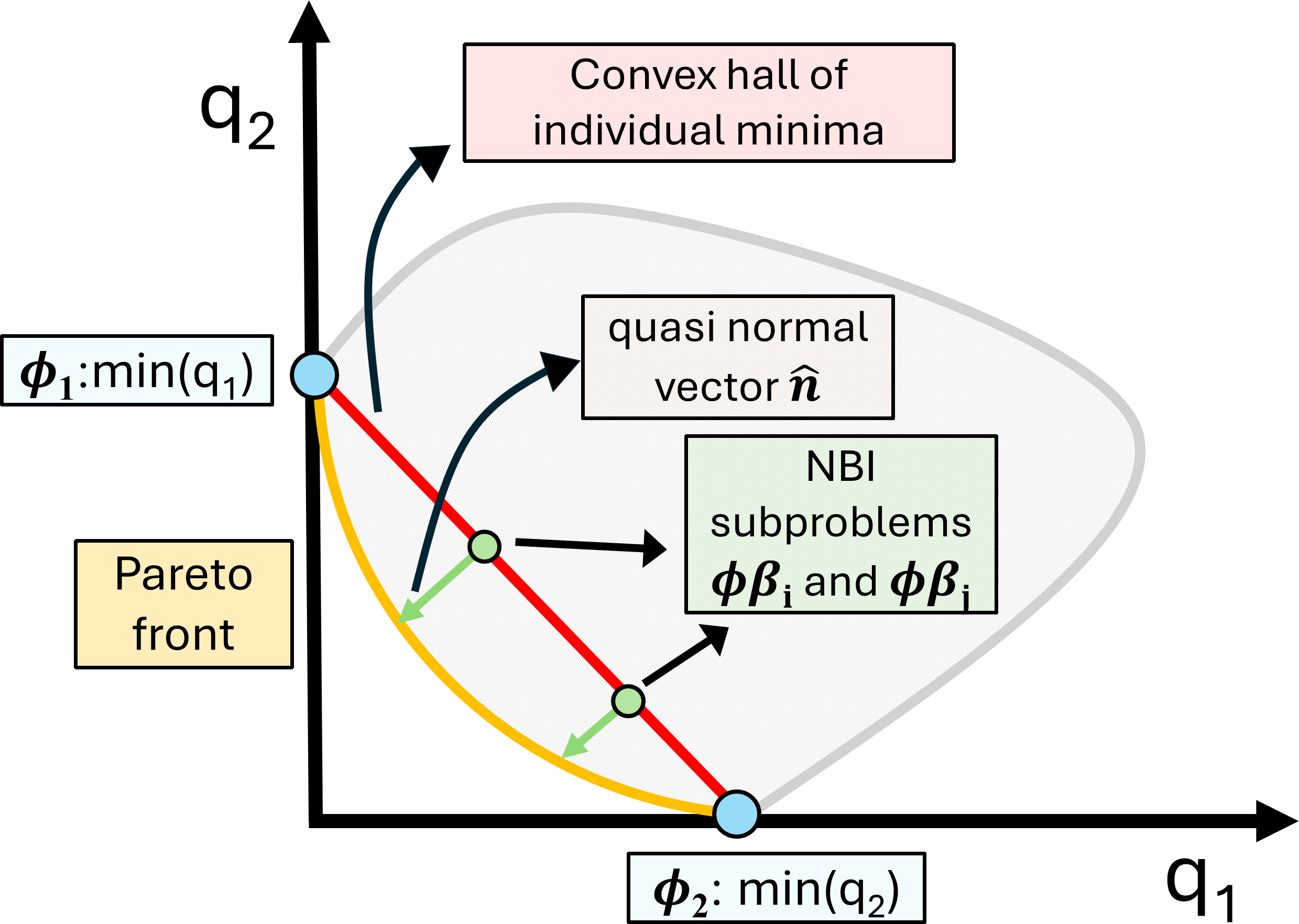}
\caption{Illustration of the normal boundary intersection method for discovering the optimal Pareto front. The method begins by initializing subproblems over the Convex Hull of Individual Minima (CHIM) and then sequentially solving them by searching along a quasi-normal vector directed toward the origin. The solution corresponding to the farthest point from the CHIM is identified as part of the Pareto front.}
\label{case_visual}
\end{figure}

Such a description of problem formulation space and the associated solutions is similar to a multi-objective optimization technique known as the Normal Boundary Intersection (NBI) method~\cite{das1997nonlinear,doi:10.1137/S1052623496307510}. As illustrated in Fig.~\ref{case_visual}, the NBI attempts to discover the entire Pareto front by distributing points along the line (or hyperplane in higher-dimensional spaces) that connects global minima of each QoI and solving each initialized subproblem to obtain one solution on the Pareto front. From this explanation, it is evident that there is a close relationship between the NBI and the problem formulation space mechanism that defines problems according to the designer's preferences. In contrast to the NBI, we are not required to solve every subproblem, but only those that a designer is interested in. However, the NBI still suggests a suitable mechanism to tackle each subproblem. In the following, we define the mathematics and terminologies behind the NBI method.

Following Refs.~\cite{das1997nonlinear,doi:10.1137/S1052623496307510}, let $\textbf{x}_i^*$ be the global minimizer of QoI $\mathbf{Q}_i (\textbf{x})$ for $i=1,...,K$. Matrix $\Phi$ is defined as a $K \times K$ matrix whose $i$th column is $\mathbf{Q}_i (\textbf{x}_i^*) - \mathbf{Q} (\textbf{x}_i^*)$. This adjustment is already applied to the example in Fig.~\ref{case_visual} showing the global minima touching the horizontal and vertical axes.
The set of points in $\mathbb{R}^K$ that are convex combinations of the columns of $\Phi$ is known as the convex hull of individual minima (CHIM). Assuming that $\hat{n}$ is the unit normal to the CHIM pointing toward the origin, a sub-problem to solve is:
\begin{equation}
\label{sub}
\begin{aligned}
   &\max_{\textbf{x},c} c\quad
    \textrm{s.t.}\quad \Phi \beta + c\hat{n} = \mathbf{Q}(\textbf{x}),\\
    &\quad h(\textbf{x})=0,\;g(\textbf{x})\leq0,\quad \textbf{x} \in \mathcal{X}.
\end{aligned}
\end{equation}
Here, $\beta$ is a vector of coefficients used to linearly combine the columns of $\Phi$ to determine a particular location on the CHIM. In the 2D example of Fig.~\ref{case_visual}, the red line represents the CHIM, connecting the global minimizers of each QoI. NBI subproblems as points on the CHIM are generated by varying $\beta$. Therefore, $\beta$ represents the space of all definable subproblems concerning different QoIs, which we refer to as the problem formulation space. To ensure the optimal Pareto front is discovered with a good distribution of solutions, we define the quasi-normal vector $\hat{n}$ as the summation of the columns of $\Phi$, which allows us to search both sides of the CHIM and guarantees a uniform distribution of solutions on the Pareto front.

A generic problem from the problem formulation space can be represented by a subset of QoIs and their respective parameters:
\begin{equation}
    p_i = (\mathcal{Q}_i,\mathcal{B}_i).
\end{equation}
Note that when using the NBI method, the parameters are the coefficients used to locate a specific subproblem on the CHIM, as in Eq.~\ref{sub}. The dissimilarity between two problems can be determined by a distance metric, defined as the solution to an optimal transport problem~\cite{vershik2013long,villani2009optimal}:
\begin{equation}
    d(p_i,p_j) = ||\mathcal{B}_i-\mathcal{B}_j||_1,
\end{equation}
which represents the minimum cost of moving from one problem to another. This definition lets us consider the problem formulation space as a metric space $(P,d)$, allowing us to map from the problem formulation space to a related attribute, such as a utility value.

In this work, we construct a Gaussian process (GP) over the problem formulation space to map each problem to a utility value. A GP is a powerful statistical tool for probabilistic modeling in the Bayesian context. Generally, a GP is a random process defined on a domain characterized by a mean and a covariance function~\cite{Wagner2023SemiAutonomousPF}. Given data $\mathcal{D}_n=\{(p_i,v_i)\}_{i=1}^n$, with $v_i=U(p_i)$,
\begin{equation}
\label{GP}
\begin{aligned}
\mathbb{E}[\hat{v}(p_*)] &= \mathbf{k}_*^\textrm{T}K^{-1} \mathbf{v},\\
\mathbb{V}[\hat{v}(p_*)]  &=  k(p_*,p_*) - \mathbf{k}_*^\textrm{T}K^{-1} \mathbf{k}_*,
\end{aligned}
\end{equation}
where $\mathbf{v}=[v_1,v_2,\dots,v_n]$, $k$ is a real-valued kernel function over the problem formulation space, $K$ is an $n \times n$ matrix with $(i,j)$ entry $k(p_i,p_j)$, $k_*$ is a column vector of size $n$ with its $i^\textrm{th}$ entry given by $k(p_*,p_i)$, and $U(p_i)$ is a multi-attribute utility function. A common choice for the kernel is the squared exponential covariance function:
\begin{equation}
\label{kernel}
k(p_i,p_j) = \exp\left(\frac{-d(p_i,p_j)^2}{l^2}\right),
\end{equation}
where $l$ is the characteristic length-scale indicating the correlation strength between points (problems, in our case) in the space.

Given a GP constructed over the problem formulation space using data collected to date, Bayesian optimization determines the next best problem to solve by finding the maximizer of an acquisition function, such as Expected Improvement (EI) or Knowledge Gradient (KG). The ultimate goal is to discover the optimal problem formulation in $\beta$ space that maximizes a multi-attribute utility function incorporating the latent preferences of the decision-maker. Building such utility functions is challenging since interactions among multiple QoIs may not be fully comprehensible. A potential solution is to conduct A/B testing, asking the decision-maker to vote for preferred designs given the designs and respective properties of previously solved problems. Accordingly, a value is assigned to each problem based on the number of A/B test wins. Iteratively, the decision-maker may update preferences while the framework explores other regions of the problem formulation space, exploiting information acquired from previously solved problems without restarting the discovery campaign. In other words, instead of defining a specific design problem, the decision-maker guides the search toward an optimal problem formulation by inputting their preferences without explicitly formulating a design problem. For demonstration purposes, we keep humans out of the loop and build a simple multi-attribute function to perform optimization over the problem formulation space.

\subsection{AI-assisted Problem Discovery}
\begin{figure*}[htb!]
\centering
\includegraphics[width=\textwidth]{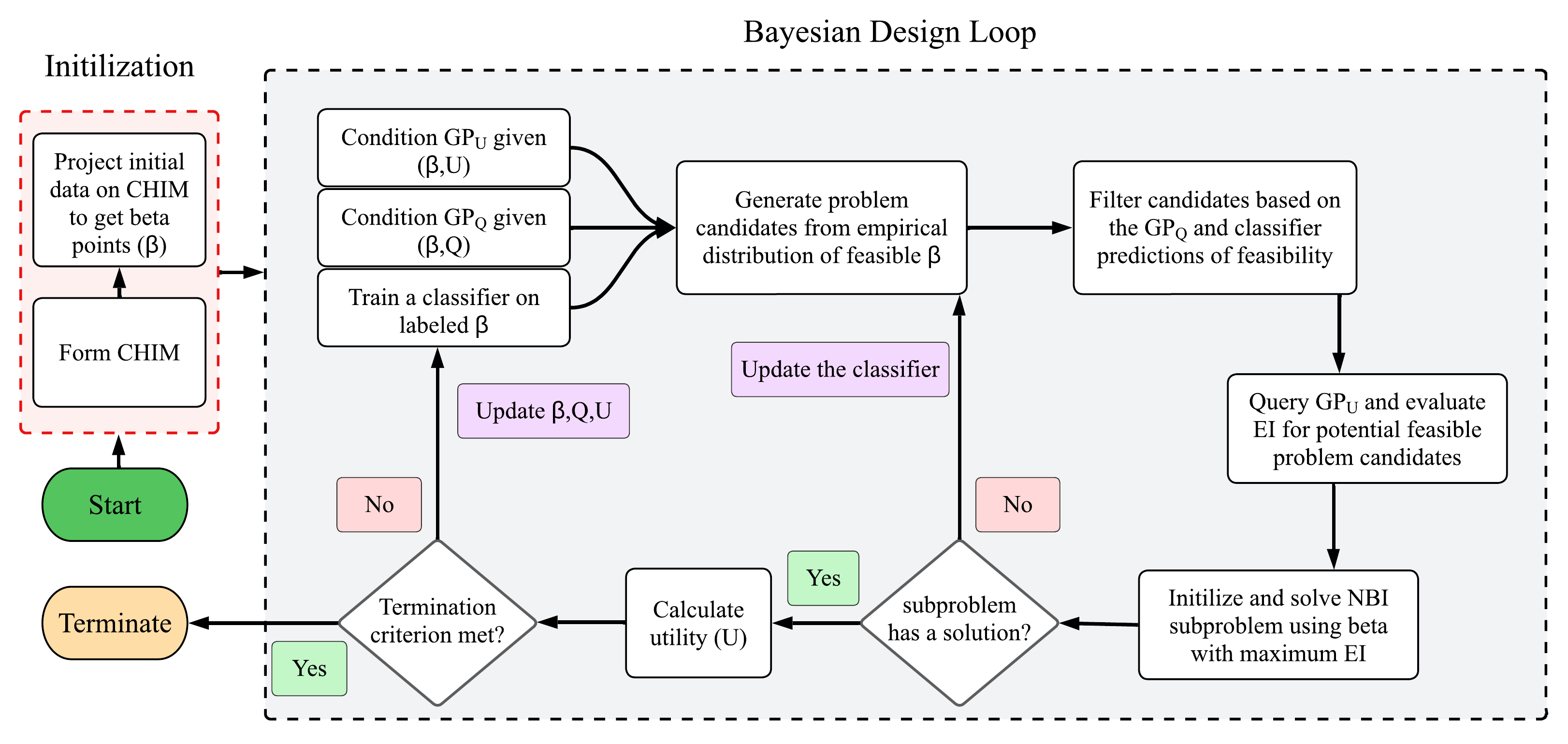}
\caption{Step-by-step representation of our proposed framework. After completing the initialization steps, the Bayesian design loop explores the problem formulation space to discover the optimal problem whose solution maximizes a utility value function.}
\label{diagram}
\end{figure*}

Our AI-assisted discovery loop---illustrated in the flowchart in Fig.~\ref{diagram}---begins by initializing the design space using expert estimates to form the Convex Hull of Individual Minima (CHIM) and projecting initial training data onto this hull to obtain $\beta$ vectors. Next, a GP surrogate (GP\textsubscript{U}) maps these problem formulations to utility values, while a paired regression model and classifier (GP\textsubscript{Q}) filter out infeasible formulations. Candidate problem formulations are generated via kernel density estimation, evaluated using an acquisition function (e.g., Expected Improvement), and the best candidate is selected for solution. The loop iteratively updates the surrogate models with new data and incorporates decision-maker feedback (via A/B testing) until a termination criterion is met.

In the following, we describe each step in more detail, the challenges associated with optimization over the problem formulation space using the NBI mechanism, and how our algorithm addresses these challenges.

The initialization step is completed by forming the CHIM based on expert estimates of the global minima of QoIs to start the Bayesian design loop. In contrast to the original implementation of the NBI~\cite{das1997nonlinear,doi:10.1137/S1052623496307510}, we propose not to bound $\beta$ space to a unit hypercube, and we let the framework search for negative regions while holding the $||\beta_i||_1=1$ constraint. This approach is motivated by two observations: bounded $\beta$ space cannot represent the projection of the entire objective space on the CHIM, particularly in higher-dimensional problems, and it relaxes the need for exact global minimizers. Since both negative and positive regions are searched, discovering a better minimum does not significantly impact performance due to the CHIM being formed from rough estimations.

At the beginning of a discovery campaign, previously evaluated designs can serve as initial training data. The second initialization step projects properties from the objective space onto the CHIM using linear algebra to obtain the associated $\beta$ vectors. In this way, the problem formulations corresponding to the initial training data are identified, allowing us to relate problem formulations to a utility value.

The first step in the Bayesian design loop is to construct surrogate models and classifiers to guide the search. The first surrogate, GP\textsubscript{U}, maps problem formulations ($\beta_i$) to their respective utility values. This GP provides probabilistic predictions on the utility of unsolved problem formulations using an acquisition function (in our case, Expected Improvement).

One challenging issue is that not every problem formulation may yield a solution. As illustrated in Eq.~\ref{sub} and Fig.~\ref{case_visual}, the NBI method seeks the solution to a subproblem (i.e., a particular problem formulation in our case) along the quasi-normal vector $\hat{n}$. However, in three or more dimensions, there is no guarantee that this line will intersect the objective space; thus, a solution may not exist. In the original implementation of the NBI method~\cite{das1997nonlinear,doi:10.1137/S1052623496307510}, the authors noted that some subproblems fail to produce a solution. To the best of our knowledge, subsequent studies have largely neglected this issue without further analysis.

Given our focus on developing a highly efficient design approach under budget constraints, it is crucial to determine whether a given problem formulation is feasible before further processing. To address this, we construct a regression model and a classifier to identify feasible problem formulations. Specifically, our GP\textsubscript{Q} maps problem formulations from $\beta$ space to the corresponding properties. If a formulation is feasible, the intersection between the quasi-normal line and the objective space should lie within the range of acceptable values for each QoI; otherwise, no solution exists. GP\textsubscript{Q} is trained on data from problem formulations that yield solutions to predict the intersection point and assess feasibility. We then pair GP\textsubscript{Q} with a binary classifier, which is updated with labeled data: if a proposed formulation is deemed infeasible, it is flagged accordingly. This two-step filtering process reduces the number of failed candidate formulations before they enter the Bayesian design loop.

It is important to note that the issue of infeasible problem formulations arises only when working in $\beta$ space, particularly in fully computational settings. In contrast, in real experimental scenarios with a human in the loop, utility values are assigned directly to design points based on decision-maker votes or rankings, making precise problem formulation unnecessary.

The next step is to generate problem formulation candidates to evaluate potential improvements in utility by solving a new problem. Since we are not bounding the $\beta$ space, conventional space-filling techniques such as Latin hypercube sampling are not suitable. Instead, we sample from the distribution of feasible points in $\beta$ space using its kernel density estimate. This method obviates the need for exact bounds and generates more problem formulations from regions with a higher probability of yielding a solution. The generated samples are then filtered using GP\textsubscript{Q} and the classifier. The remaining samples are queried from GP\textsubscript{U} to compute EI and determine the best candidate formulation to solve. If a solution is not found, the classifier is updated with the new data point, and the process is repeated until success. Finally, the utility is calculated, and all datasets and surrogate models are updated accordingly. The Bayesian design loop repeats until a termination criterion, such as budget or time limits, is met.

The most important part of the algorithm is constructing a utility function to evaluate the quality of a design according to decision-maker preferences. Due to the limited human comprehension of high-dimensional data and the coupled relationships among subsets of such spaces, mathematical representations are challenging to create. However, A/B testing enables the comparison of different designs and their associated properties, or even ranking them based on decision-maker preferences. A machine can discern the latent preferences of the decision-maker and suggest a problem formulation that aligns with their expectations for a good design at any stage. A simple approach is to assign scores to designs based on votes or rankings and map problem formulations to the calculated utilities using a probabilistic model such as a GP.

\section{Demonstration Problem: Optimal Problem Formulation for Alloy Development}
\label{demonstration}

\subsection{Refractory Alloys for Next-Generation Turbine Blades}
The design of materials for gas turbine engine (GTE) blades is especially challenging due to the need to simultaneously balance multiple objectives and constraints. Key properties---including ductility, yield strength, solidification range, and density---often interact in conflicting ways, further complicating the design process. The vast chemical space and conflicting property requirements necessitate the use of advanced machine learning and active learning techniques to accelerate alloy discovery and optimization.

To benchmark our method, we propose an \emph{in silico} synthetic design problem using the Mo-Nb-Ti-V-W element system. In this benchmark, candidate alloys must meet several critical criteria: they must exhibit room-temperature ductility, maintain high yield strength at elevated temperatures, have a narrow solidification range, and possess low density. The threshold values and property specifications are derived from real-world alloy design campaigns~\cite{arpa-e_2020,acemi2024multi}.

In this \emph{in silico} example, various data sources are leveraged to predict the ground-truth properties of candidate alloys. Ductility is estimated using Cauchy pressure, which indicates the alloy's ability to deform without failure. Yield strength is predicted through the Curtin-Maresca model~\cite{MARESCA2020235}, a well-established model for assessing the strength of high-entropy alloys. Additionally, solidification range and density are computed using Thermo-Calc equilibrium and property modules. Together, these models provide a comprehensive understanding of the alloys' performance, allowing for the rapid identification of optimal candidates for further investigation.

For demonstration purposes, we simulate a static A/B testing scenario by constructing a multi-attribute utility function that encodes fixed preferences and thresholds on multiple QoIs relevant to the alloy performance requirements of gas turbines in power generation. This approach is analogous to static A/B testing, where the preferences remain unchanged over time.

Using the NBI approach to solve subproblems within a multi-objective optimization framework, we define the problem formulation space and implement Bayesian optimization to efficiently discover the formulation that maximizes the utility value. Recognizing that not all formulations yield feasible solutions, we train a classifier and a Gaussian process (GP) to filter the search space before further evaluation. The final results reveal a “sweet spot” that meets all design thresholds and best aligns with the decision-maker’s preferences, since any further improvement in a given QoI would lead to unacceptable degradation in others, thereby reducing the overall utility.

\subsection{Approximating Expert Preferences with Utility Functions}
For demonstration purposes, instead of iteratively asking a human decision-maker to rank alloys or conduct A/B tests, we simulate a static scenario by encoding the preferences and thresholds into simple utility functions. Here, ``static'' means that the preferences do not change over time as the utility functions remain fixed.

We then aim to discover the optimal problem formulation that yields properties addressing the trade-offs and constraints while satisfying the decision-maker's preferences. In this scenario, we do not simply minimize or maximize the quantities of interest; instead, we map each QoI value to a corresponding utility. Any constraint can also be treated as a QoI, with a non-zero utility specifying the desired range.

The utility functions for this \emph{in silico} problem are illustrated in Figure~\ref{utils}. These functions were specified based on expert opinion and institutional knowledge acquired from several experimental alloy design campaigns~\cite{acemi2024multi,hastings2024interoperable,mulukutla2024illustrating}.

For ductility, the utility derived from Cauchy pressure is modeled as an exponential-linear piecewise function. Initially, incremental increases in ductility yield significant boosts in utility, reflecting the critical need to enhance ductility at lower levels. Once the Cauchy pressure reaches 70 GPa, the utility continues to improve linearly with additional increases, but with diminishing returns. This approach captures the desire to maximize ductility while recognizing that beyond a certain point, further improvements are less valuable.

The utility function for yield strength follows a sigmoid model, with a critical point at 200 MPa where 99\% of the total utility is achieved. Beyond this threshold, the utility plateaus, indicating that further increases offer no substantial additional benefit. This reflects a strong preference for achieving sufficient yield strength while acknowledging that improvements beyond this point do not significantly enhance performance.

For density, a sigmoid deactivation function is used, with an inflection point at 9 g/cc where half the utility is achieved. This function emphasizes reducing density below 10 g/cc, with diminishing returns as density falls below 9 g/cc, and no additional utility gained by reducing density below 8 g/cc.

Finally, the solidification range is represented by a linearly decreasing utility function, prioritizing alloys with a narrow range to minimize cracking risks.

The minimum and maximum values for these utility functions are defined based on the property ranges observed within the design space. Using a full-factorial query of the Mo-Nb-Ti-V-W alloy space at 5 atomic percent, the extremas of these properties were used to bound the utility functions.

The multi-attribute utility function for each experiment is given by:
\begin{equation}
    U = 1.5 \times u_{cp} + 1.3 \times u_{ys} + u_{density} + u_{sr},
\end{equation}
where the weights reflect the relative importance assigned in real alloy design campaigns~\cite{arpa-e_2020,acemi2024multi,hastings2024interoperable,mulukutla2024illustrating}.

\subsection{Results}
The campaign was initialized using 40 known alloys and their properties to construct the surrogate models and classifiers. Since the discovery is performed over the problem formulation space, $\beta$, the classifier and regression model help identify formulations that yield a solution. The classifier predicts feasibility based on labeled training data, while the regression model (GP\textsubscript{Q}) predicts the intersection of the quasi-normal vector with the QoIs to ensure it lies within acceptable ranges. After forming the CHIM, 40 data points are projected onto the CHIM to obtain the respective problem formulations for constructing GP\textsubscript{U}, GP\textsubscript{Q}, and the classifier.

\begin{figure}[htb!]
\centering
\includegraphics[width=\columnwidth]{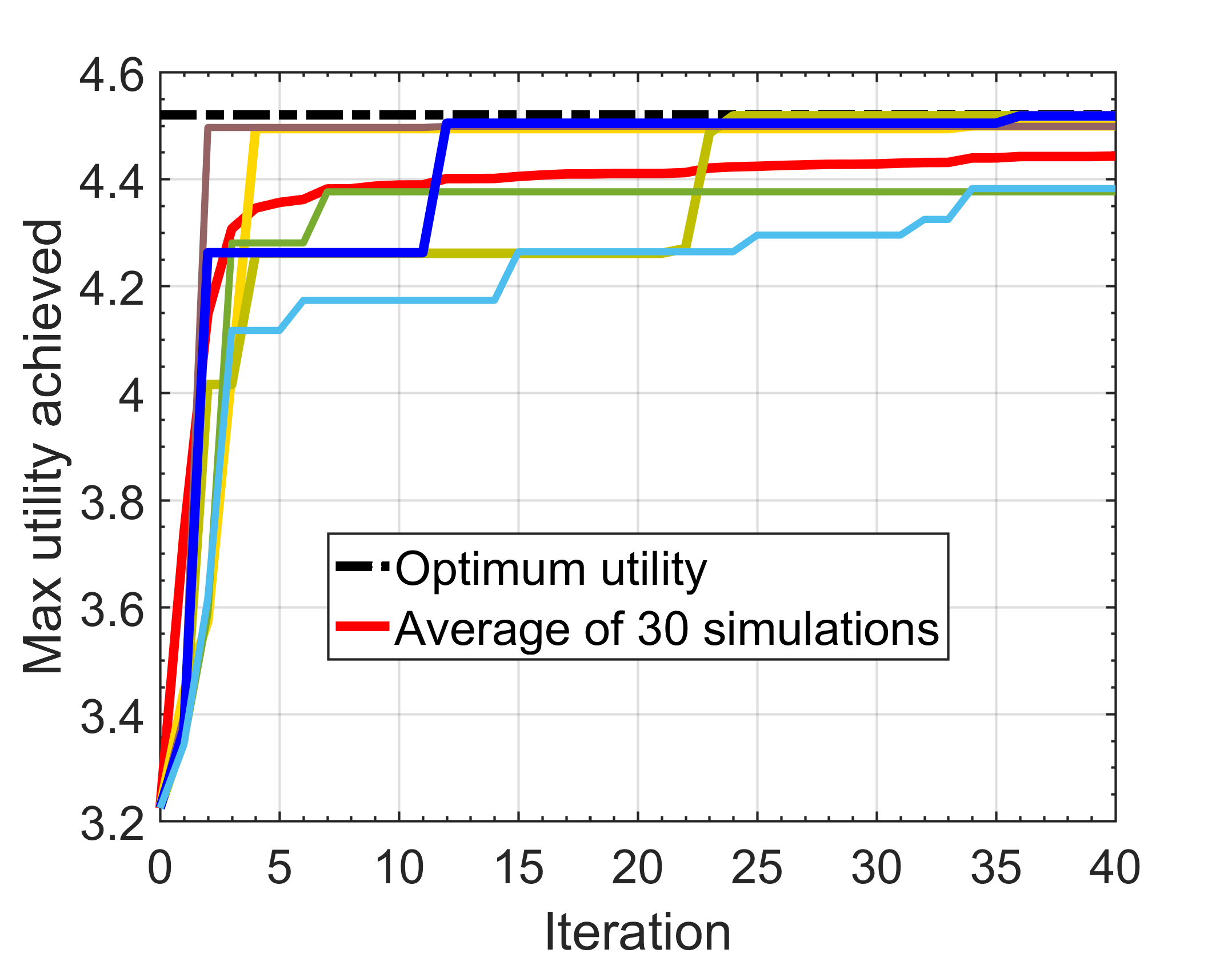}
\caption{Comparison of optimum utility versus the average achieved utility from 30 simulations. Best and worst simulation results are also shown to illustrate variation.}
\label{average}
\end{figure}

We repeated the simulation 30 times for 40 iterations to obtain average performance metrics. Figure~\ref{average} shows the average utility achieved over 30 replications, along with some of the best and worst performing simulations, compared to the maximum achievable utility. The first few iterations in each simulation yield significant utility improvements by quickly recognizing near-optimal formulations. Despite the stochastic nature of Bayesian optimization, major improvements are observed early on.

\begin{figure}[htb!]
\centering
\includegraphics[width=\columnwidth]{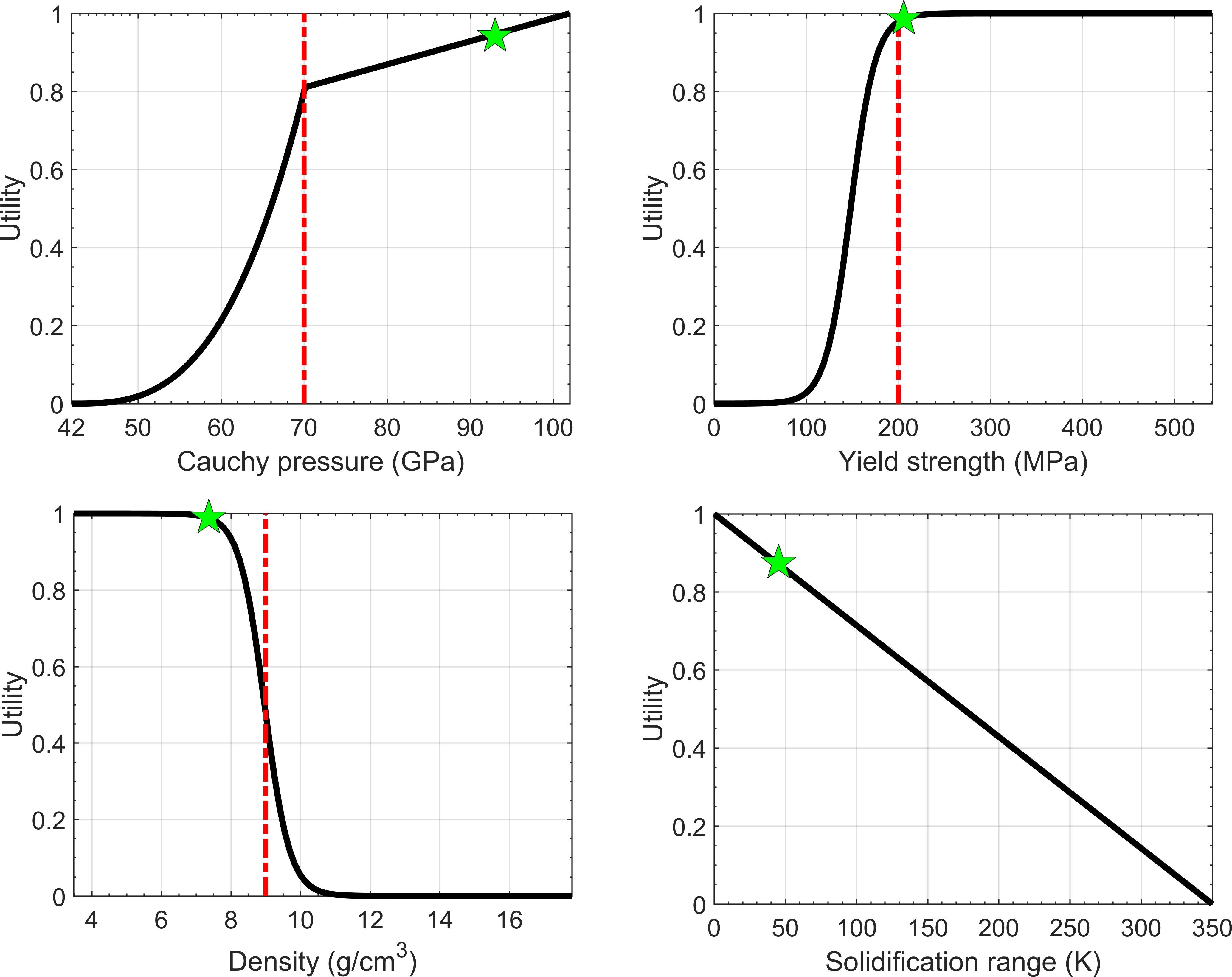}
\caption{Utility value functions constructed to capture the decision-maker's preferences for different QoIs. Green stars indicate the optimum properties that maximize the multi-attribute utility function.}
\label{utils}
\end{figure}

Figure~\ref{utils} illustrates the utility functions, thresholds for each QoI, and the discovered optimal values that maximize the multi-attribute utility function. For example, the yield strength is near the threshold, suggesting that further increases would detrimentally affect ductility and density, thereby lowering overall utility. The discovered optimal solution represents a sweet spot where trade-offs between QoIs are balanced to best satisfy decision-maker preferences.

Tables~\ref{Properties} and~\ref{chemistry} summarize the optimal alloy's constituent elements and corresponding properties, which together maximize the multi-attribute utility function.


\begin{table}[htb!]
\centering
\caption{Optimal property values yielding the highest utility. Here, $CP$ denotes Cauchy pressure, $\sigma_y$ denotes yield strength, $\rho$ denotes density, and $\Delta T$ denotes the solidification range.}
\label{Properties}
\resizebox{\columnwidth}{!}{%
\begin{tabular}{ccccc}
\hline
Property & $CP$ & $\sigma_y$ & $\rho$ & $\Delta T$ \\ \hline
Optimal  & 91.4 GPa      & 206.6 MPa  & 7.8 g/cc & 44.7 K   \\ \hline
\end{tabular}%
}
\end{table}

\begin{table}[htb!]
\centering
\caption{Weight percentages of each element in the optimal alloy.}
\label{chemistry}
\begin{tabular}{cccccc}
\hline
Element   & Mo    & Nb   & Ti   & V     & W    \\ \hline
Weight \% & 33.99 & 2.05 & 0.16 & 63.80 & 0.00 \\ \hline
\end{tabular}
\end{table}

\begin{figure}[htb!]
\centering
\includegraphics[width=\columnwidth]{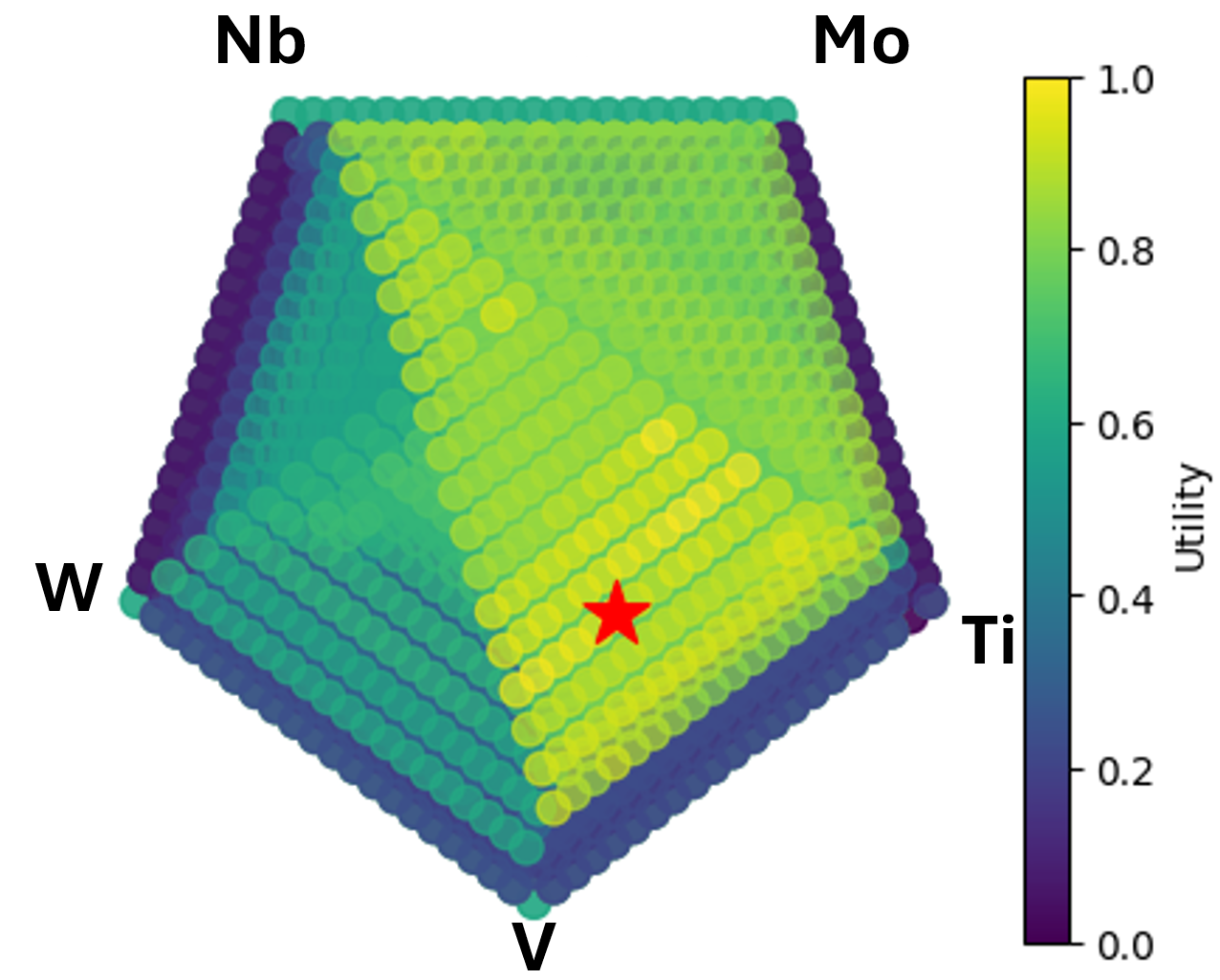}
\caption{Affine projection of the Mo-Nb-Ti-V-W composition space into 2D. The vertices represent pure elements, and the color scale indicates the normalized utility values.}
\label{umap}
\end{figure}

To better understand the variation of the utility function with alloy chemistry, we project the Mo-Nb-Ti-V-W composition space into 2D using an affine projection. In this projection, each alloy composition is represented within a regular pentagon, with vertices corresponding to the pure elements. Compositions enriched in Mo are closer to the Mo vertex, while higher-entropy alloys cluster near the center. The utility function is normalized based on the maximum achievable value and visualized using a color scale (see Fig.~\ref{umap}).

\begin{figure}[htb!]
\centering
\includegraphics[width=\columnwidth]{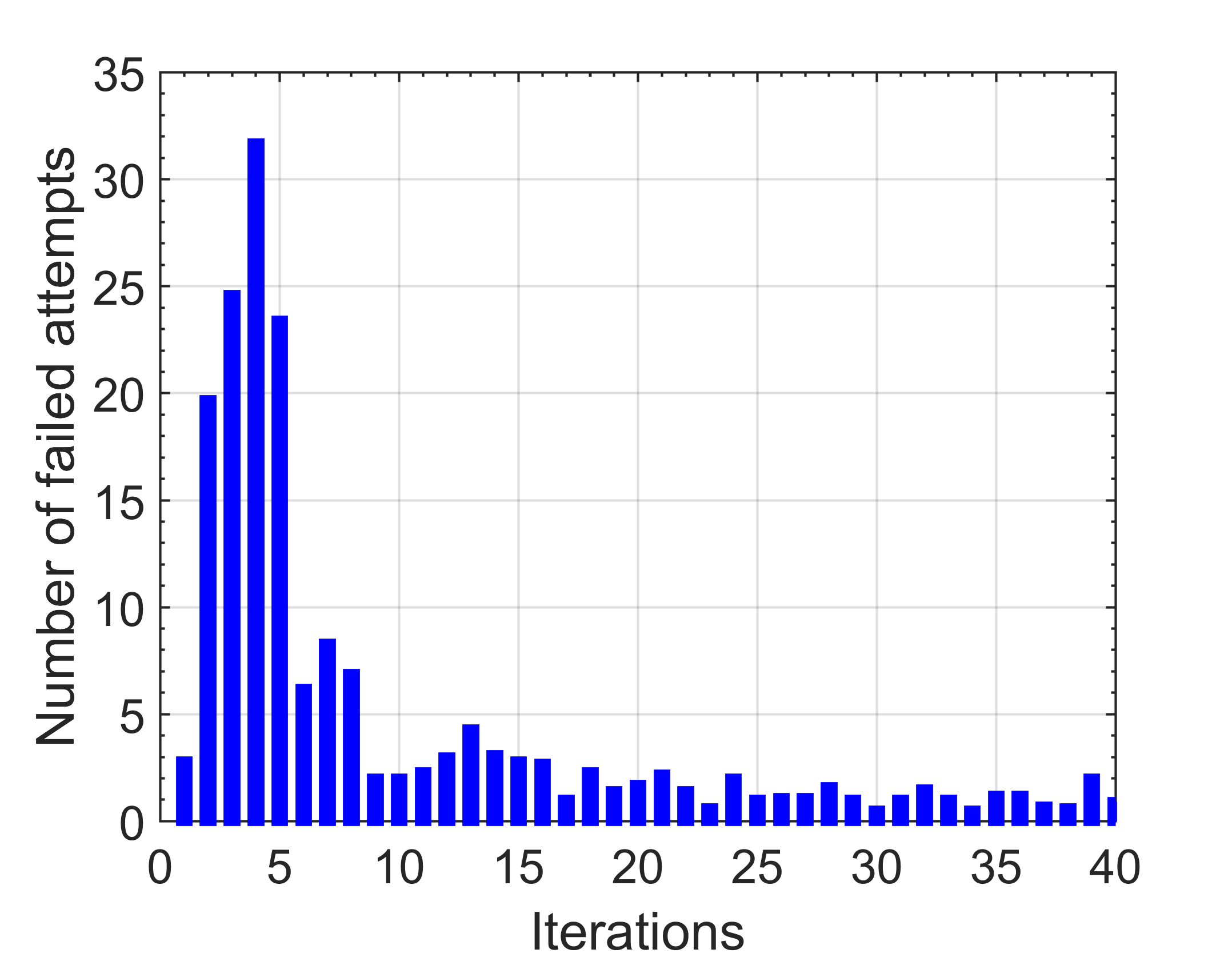}
\caption{Average number of failed attempts from 30 simulations. Early iterations have many false predictions; incorporating these observations improves the classifier's performance in later iterations.}
\label{fails}
\end{figure}

In contrast to human-in-the-loop scenarios—where Bayesian optimization directly explores the alloy space to propose new experiments—in our demonstration, Bayesian optimization is executed over the problem formulation space $\beta$ in a fully computational setting. Consequently, handling formulations that yield no solution is critical. We address this by filtering candidate formulations using a classifier and regression model. Initially, with only 40 training datasets, the classifier has not encountered any infeasible formulations, so false predictions are expected early on. As more observations are collected, the system learns and improves its filtering capability. Figure~\ref{fails} plots the average number of false predictions that passed both the classifier and regression model versus iteration. Early iterations show many failed attempts, but incorporating this feedback enhances performance in later iterations.

\begin{figure}[htb!]
\centering
\includegraphics[width=\columnwidth]{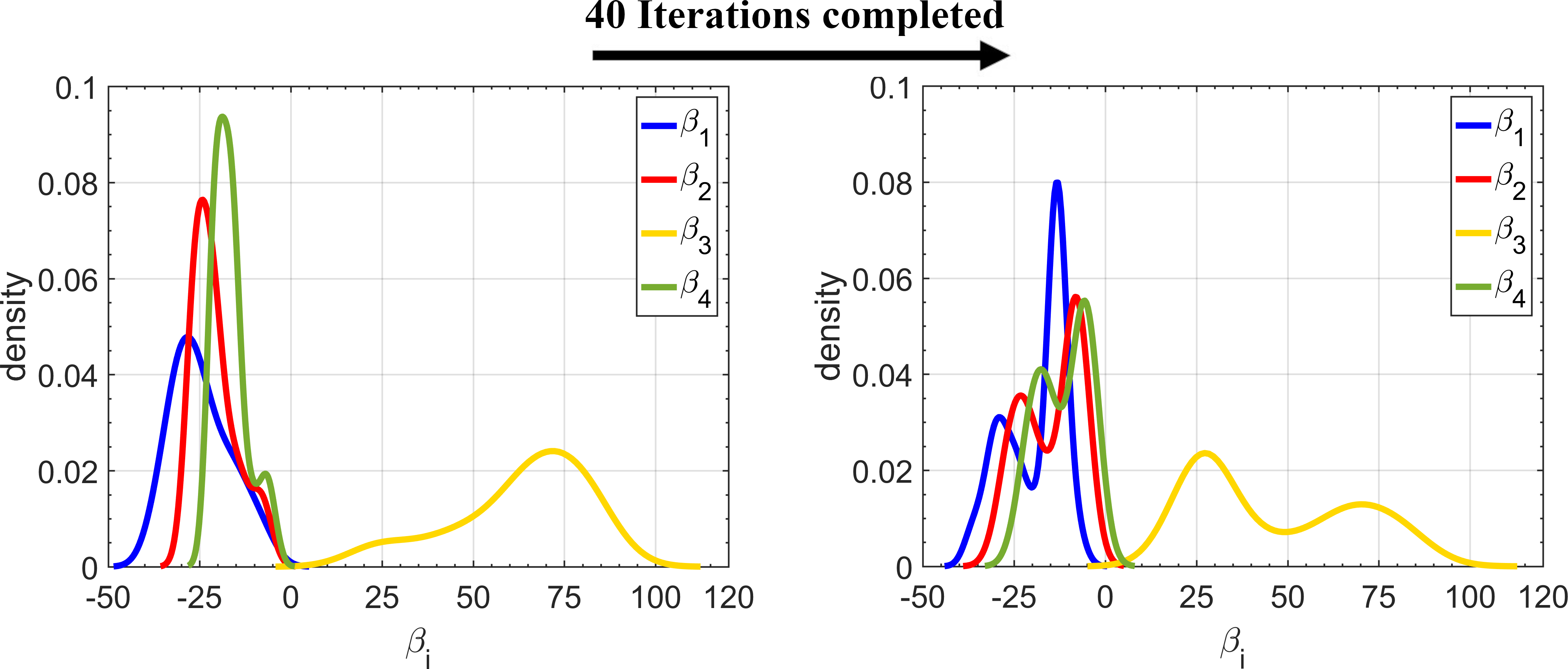}
\caption{Kernel density estimates of the distribution of feasible problem formulations from 40 initial training data and after 40 iterations. New peaks correspond to the optimal problem formulation region.}
\label{dists}
\end{figure}

Since we are not bounding $\beta$ to a unit hypercube as in the original NBI, we generate new candidates each iteration by sampling from the kernel density estimation of the distribution of feasible problem formulations and iteratively updating it with new observations. Figure~\ref{dists} shows the initial and final kernel density estimates of the distribution of $\beta$ values. New peaks in the distribution correspond to the optimal formulation region discovered by the framework, demonstrating the importance of searching $\beta$ space beyond a unit hypercube.

\section{Conclusions and Future Work}
\label{conclusions}

Autonomous design and experimentation require mechanisms that enable machines to identify the most valuable problem to solve at each stage of a discovery task. In this study, we introduce a novel approach that employs Bayesian optimization over the problem formulation space to search for the optimal formulation in line with the decision-maker’s preferences. The system learns these preferences from tabulated design data and through pairwise comparisons or rankings of observed design points, allowing decision-makers to update their criteria dynamically as the design process unfolds—thereby eliminating the need to precisely define the design problem at the outset.

For demonstration purposes, we simulated a scenario by constructing a multi-attribute utility function that encodes fixed preferences and thresholds on multiple QoIs relevant to the alloy performance requirements of gas turbines in power generation. This approach is analogous to static A/B testing, where the preferences remain unchanged over time.

Using the NBI approach to solve subproblems within a multi-objective optimization framework, we defined the problem formulation space and implemented Bayesian optimization to efficiently discover the formulation that maximizes the utility value. Recognizing that not all formulations yield feasible solutions, we trained a classifier and a Gaussian process (GP) to filter the search space before further evaluation. The final results revealed a “sweet spot” that meets all design thresholds and best aligns with the decision-maker’s preferences, since any further improvement in a given QoI would lead to unacceptable degradation in others, thereby reducing the overall utility.

Future work will integrate human feedback into the loop to drive discovery in real experimental settings, allowing Bayesian optimization to explore the design space directly according to evolving preferences—thus bypassing the need for the NBI-based mathematical formulation. Moreover, while our discussion has focused on a single decision-maker’s preferences, in practice different stakeholders may have varying opinions on the optimal design. In such cases, each decision-maker can rank the designs according to their preferences, and the machine can assign a composite utility value (e.g., via a scoring system) to each design. Bayesian optimization can then propose the next experiment based on a consensus ranking, ensuring that the suggested design is well-regarded by all parties.

\section*{Acknowledgements}
We acknowledge the support of the National Science Foundation (NSF) through Grant No. 2323611 (\emph{DMREF: Optimizing Problem Formulation for PrinTable Refractory Alloys via Integrated MAterials and processing co-design (OPTIMA)}). RA and DA also acknowledge support from the Army Research Laboratory (ARL) through Cooperative Agreement Number W911NF-22-2-0106 supporting the \emph{BIRDSHOT Center} at Texas A\&M University, and under the DEVCOM \emph{High-throughput Materials Discovery for Extreme Conditions (HTMDEC)} program. RA and BV also acknowledge support from the U.S. Department of Energy (DOE) ARPA-E ULTIMATE Program through Project DE-AR0001427. BV acknowledges the support of NSF through Grant Nos. 1746932 (GRFP) and 1545403 (NRT-D3EM). Calculations were carried out on resources provided by the Texas A\&M High-Performance Research Computing (HPRC) facility.

\bibliographystyle{elsarticle-num}
\bibliography{asme2e}

\end{document}